\documentclass[prl,twocolumn,superscriptaddress,preprintnumbers,
               showpacs,amsmath,amssymb]{revtex4}

\usepackage{graphicx}
\usepackage{epsfig}
\usepackage{color}

\begin{document}

\title{Superconductivity in doped cubic silicon: an {\it ab initio} study}

\author{E. Bourgeois and X. Blase}

\address{ Laboratoire de Physique de la Mati\`ere Condens\'ee et
Nanostructures, Universit\'e  Lyon I; CNRS, UMR 5586,
Domaine Scientifique de la Doua, F-69622 Villeurbanne cedex; France.}

\begin{abstract}
We study within a first-principle approach the band structure,
vibrational modes and electron-phonon coupling in boron, aluminum
and phosphorus doped silicon in the diamond phase. Our results
provide evidences that the recently discovered superconducting
transition in boron doped cubic silicon can be explained within a
standard phonon-mediated mechanism. The importance of lattice
compression and dopant related stretching modes are emphasized. We
find that T$_C$ can be increased by one order of magnitude by
adopting aluminum doping instead of boron.
\end{abstract}



\date{\today}
\maketitle


The experimental discovery by Bustarret and coworkers
\cite{Bustarret06} of a superconducting (SC) transition in heavily
boron-doped silicon in the diamond phase (labeled {\it c}-Si in
what follows) concluded a long history of research on such a
transition in silicon-based systems and in doped semiconductors in
general.\cite{Cohen64a} Until recently, a SC behavior in silicon
had only been observed in high-pressure metallic phases such as
the hexagonal or $\beta$-tin  structures \cite{Martinez85} or
low-pressure cage-like doped clathrate phases.
\cite{Yamanaka95,Conneta03} This latter structure, which becomes
superconducting upon heavy doping, bears much similarities with
the diamond phase, as it is a semiconducting {\it sp}$^3$ network,
but with a band gap in the visible range.\cite{Conneta01}

Besides doped {\it sp}$^3$ silicon clathrates, superconductivity
in {\it c}-Si was recently made likely by the discovery of a SC
transition in highly boron-doped carbon diamond. \cite{Ekimov04}
Tunnelling spectroscopy,\cite{Sacepe} reflectivity measurements
\cite{Ortolani} and first-principles studies within the density
functional theory (DFT), performed in the virtual crystal
approximation (VCA) \cite{Boeri04,Pickett04,Ma05} or a supercell
approach, \cite{Blase04,Xiang04,Giustino06} strongly suggested
that the transition was phonon mediated. Further, the study of the
SC origin in Si-clathrates and carbon diamond led theorists to
predict the SC transition in doped {\it c}-Si within a crude
rigid-band model for doping \cite{Conneta03} and a more accurate
VCA treatment \cite{Boeri04} with emphasis on boron doping.

In this paper, we study by means of {\it ab initio} simulations
within a supercell approach the electronic, vibrational and
electron-phonon coupling properties of boron, aluminum  {\it
p}-doped and phosphorus {\it n}-doped {\it c}-Si. We find that a
standard phonon-mediated BCS approach can account for the
experimental transition temperature observed at high boron
content. We predict further that aluminum doping would allow to
increase T$_C$ by one order of magnitude, an effect ascribed in
particular to the negative effect of lattice compression on T$_C$
in the case of boron doping.

The calculations are performed within a planewave pseudopotential
implementation~\cite{Baroni} of the DFT using the PBE functional
\cite{PBE} for exchange and correlation. Ultrasoft
pseudopotentials are used with a 20 Ry (160 Ry) cutoff for the
expansion of the wavefunctions (charge density), increased to 25
Ry (200 Ry) in the case of boron-doping.  A (2x2x2) supercell
containing 16 atoms is built with one Si atom replaced by an
impurity, leading to a $\sim$6.25$\%$ doping concentration, in the
range of the estimated 5.7-8.4 $\%$ experimental value
\cite{Bustarret06} for the superconducting boron-doped samples.
The Brillouin zone (BZ) is sampled by a (5x5x5) $\bf k$-point grid
for structural relaxation and the calculation of the phonon modes
which are obtained on a (4x4x4) $\bf q$-grid ($\bf k$ and $\bf q$
will refer to electron and phonon momentum respectively). We adopt
a much finer (10x10x10) $\bf k$-point sampling for calculating the
{\bf q}-dependent electron-phonon coupling constants
$\lambda$({\bf q}). A tetrahedron extrapolation technique is used
to accelerate the summation over phonon modes in the calculation
of the coupling constant $\lambda$.

We first minimize the energy of the system with respect to the
cell size and atomic positions. As expected, the largest
relaxation occurs for boron doping with an important  contraction
of the Si-B bonds ($\sim$ 2.1 \AA), leading to a $\sim$ 1.9$\%$
contraction of unit cell lattice parameter. Experimentally, as
doped samples are constrained in the $\widehat{xy}$ plane
(parallel to the surface) by the undoped substrate, the relaxation
was shown to be anisotropic with a $\epsilon_{zz}$ contraction
along the $\hat{z}$-axis ranging from 2.5$\%$ to 3.7$\%$  from XRD
analysis.  Using a Poisson's ratio $\nu$ of 0.28, this yields an
experimental averaged lattice compression $ \epsilon_{av} =
\epsilon_{zz} (1-\nu)/(1+\nu)$ in the 1.4$\%$-2.1$\%$ range. Our
theoretical value clearly falls within these estimates. This
lattice contraction will be shown to have important consequences
on T$_C$.

\begin{figure}[!hbtp]
\begin{center}
\includegraphics*[width=8.5 cm]{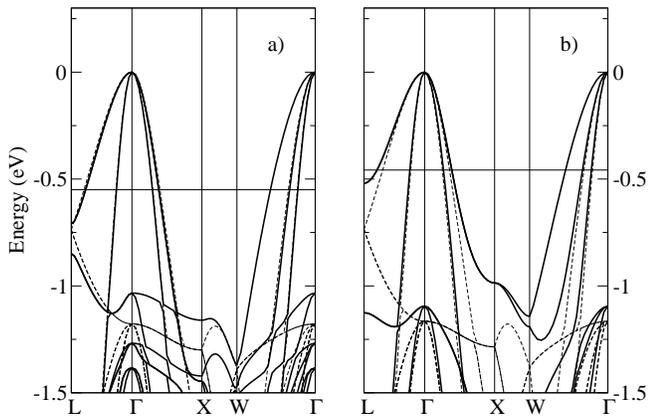}
\caption{ Electronic DFT band structures of (a) boron-, and (b)
aluminum-doped {\it c}-Si in the (2x2x2) cell. The Fermi levels
are indicated by horizontal lines. Dashed lines are for the
undoped c-Si.} \label{fig1}
\end{center}
\end{figure}

We plot in Fig.~\ref{fig1}(a) the band structure of the  B:Si cell
that we compare to that of undoped silicon at the same lattice
parameter (dashed lines). The Fermi level lies 0.55 eV below the
top of the valence bands (VBM), showing that in this limit of
large doping, Si:B is a degenerate semiconductor.  Clearly, close
to the Fermi level, the band structures of doped and pristine
silicon are very similar. In the case of {\it p}-doped aluminum
samples, the Fermi level falls 0.45 eV below the VBM
(Fig.~\ref{fig1}b).

\begin{figure}[!hbtp]
\begin{center}
\includegraphics*[width=8.5 cm]{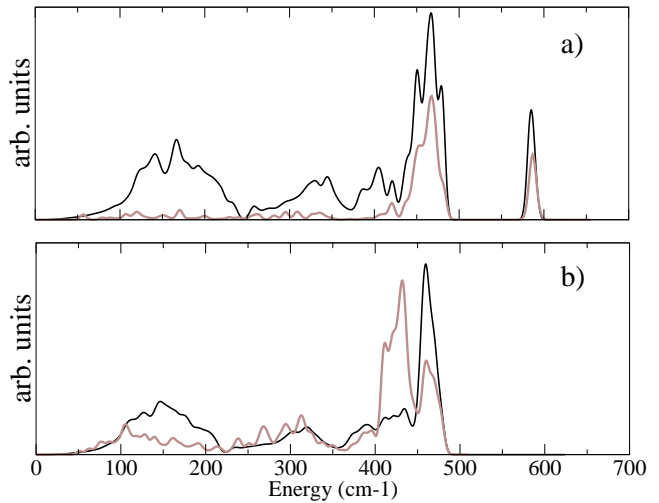}
\caption{ Phonon-DOS (black line) and Eliashberg functions (grey
thick line) for (a) boron-, and (b) aluminum-doped {\it c}-Si in
the (2x2x2) cell.} \label{fig2}
\end{center}
\end{figure}

We now turn to the vibrational properties and plot in
Fig.~\ref{fig2} the phonon density of states ({\it ph}-DOS).
 \cite{note1} As compared to the 501 cm$^{-1}$ theoretical frequency
 for the zone-center
optical modes in c-Si in the present PBE approach (520 cm$^{-1}$
experimentally), one observe a $\sim$ 40 cm$^{-1}$ softening in
B-doped and Al-doped samples. In the case of boron-doping, we find
that the effect of lattice parameter reduction shifts the
zone-center optical modes of pristine {\it c}-Si to higher
frequencies by 20 cm$^{-1}$. As a result, the effect of B-doping
really amounts to a $\sim$60 cm$^{-1}$ softening. Further, the
strong reduction of the B-Si bonds yields a well separated B-Si
(3-fold) stretching mode located $\sim$100 cm$^{-1}$ above the
optical modes continuum, a signature in excellent agreement with
the Raman peak at 590-600 cm$^{-1}$ observed experimentally in the
superconducting  B-doped samples. \cite{Bustarret06} This mode has
been already  observed experimentally in  non-superconducting
samples with a large frequency shift  under uniaxial pressure
related to the strong local stress around the dopant.
\cite{Cardona}

The Eliashberg function  ${\alpha^2}F({\omega})$, which allows to
analyze which phonons contribute to the electron-phonon coupling
constant $\lambda$, is represented in Fig.~\ref{fig2}(a-b) (grey
thick line) together with the {\it ph}-DOS (black line).
${\alpha^2}F({\omega})$ can be calculated from the knowledge of
the electron-phonon coupling matrix elements, \cite{Allen72}
namely:

\begin{eqnarray}
  {\alpha^2}F({\omega}) &=&
  N(E_F) \sum_{{\bf q}\nu} \langle | g_{{\bf q}\nu} |^2 \rangle
  \delta(\omega - \omega_{{\bf q}\nu})  \\
   \langle | g_{{\bf q}\nu} |^2 \rangle  &=&
     \int {d^3{\bf k} \over \Omega_{BZ}}
     | g_{{\bf q}\nu}^{{\bf k}nn'}|^2 \;
   { \delta({\epsilon_{n{\bf k}}}) \delta({\epsilon_{n'{\bf k+q}}})
     \over
     N(E_F)^2 }\\
 g_{{\bf q}\nu}^{{\bf k}nn'} &=&
\left({\hbar \over 2M\omega_{{\bf q}\nu} } \right)^{1 \over 2}
   \langle {\psi}^0_{n{\bf k}} | {\hat {\epsilon}}_{{\bf q}\nu} \cdot
   {\delta V^{scf} \over \delta {\hat u}_{{\bf q}\nu}}
    | {\psi}^0_{n'{\bf k+q}} \rangle
\end{eqnarray}

\noindent To accelerate the convergence with respect to phonon
momentum {\bf q}-sampling,  we observe that the largest variations
in $\langle | g_{{\bf q}\nu} |^2 \rangle$ comes from the two
$\delta$-functions in Eq. 2. By defining the nesting factor
n(${\bf q}$), that can be obtained by setting $g_{{\bf q}\nu}$ to
unity in Eq. 2, we find that $\langle | g_{{\bf q}\nu} |^2 \rangle
/ n({\bf q})$ is  smoothly varying (as a function of ${\bf q}$)
and well behaved at zone-center. From the explicit calculation of
such a ratio on the (4x4x4) {\bf q}-grid, we extrapolate $\langle
| g_{{\bf q}\nu} |^2 \rangle / n({\bf q})$
 using tetrahedron techniques on a much
finer (10x10x10)-grid in order to obtained a well converged
${\alpha^2}F({\omega})$ function. \cite{tubecase}

The integration of the Eliashberg function yields the coupling
constant $\lambda= 2 \int d\omega \;
{\alpha^2}F({\omega})/{\omega}$ with values of 0.26 and 0.36 for
the B-doped  and Al-doped systems respectively. \cite{note2} Our
$\lambda$=0.26 for B-doped {\it c}-Si is close to the 0.30 value
found in Ref.~\onlinecite{Boeri04} using the VCA for a slightly
smaller 5$\%$ doping level. In the present supercell approach, the
contribution of the B-related stretching modes at high energy
amounts to 15 $\%$ of the coupling constant. In the case of
Al-doping, there is an enhanced contribution from modes located
$\sim$ 50 cm$^{-1}$ below the softened optical modes and showing a
significant weight onto the Al displacements. This is reminiscent
of the possible importance of the B-related vibrational modes in
the case of B:C samples as suggested experimentally
\cite{Ortolani} and by theory. \cite{Blase04,Xiang04,Giustino06}

Using the standard McMillan formula, or a modified relation
accounting for the $\delta$-like shape of the Eliashberg function
(see Ref.~\onlinecite{Boeri04}), and with an average phonon
frequency $\hbar \omega_{0}$ set to the value that maximizes
${\alpha^2}F({\omega})$ ($\sim$ 470 cm$^{-1}$), one obtains for
B:Si a transition temperature T$_C$ evolving from 0.24~K to 0.06~K
for $\mu^{\ast}$ in the 0.08-0.12 range. Such a temperature is
clearly consistent with the set of values obtained in
Ref.~\onlinecite{Bustarret06} for the three superconducting
samples, with a largest T$_C$ of 0.34~K for the sample with the
highest boron content ($\sim$ 8.4 $\%$). \cite{Bustarret06}

The larger value of $\lambda$ in Al:Si can be partly attributed to
a negative pressure effect.  In the case of the Si clathrates, it
was shown experimentally and by theory \cite{Conneta03} that T$_C$
decreases under applied pressure with a $dT_C/d{\epsilon}
\sim$2~K. As a computational experiment, we study Al:Si at the
B:Si theoretical lattice constant, that is we apply a 2.1$\%$
lattice contraction. We find that $\lambda$ decreases by $\sim$
10$\%$, an effect that can be ascribed in particular to a
$\sim$6.5$\%$ density of states reduction under broadening of the
bands by applied pressure. Further, the expression of $\lambda$ as
a function of ${\alpha^2}F({\omega})$, and the inspection of
Eqs.(1-3), reveal a $(1/{\omega_{{\bf q}\nu}})^2$ dependence of a
given phonon mode to the coupling constant. This scaling reduces
by a factor $\sim (590/470)^2 \sim 1.58$ the contribution of the
three optical bands which, upon introduction of boron, are shifted
to a much higher energy, amounting to further reducing $\lambda$
by 0.03-0.04.

Changes in $\lambda$ do not affect T$_C$ directly as the average
phonon frequency prefactor $\omega_{0}$ and the Coulomb repulsion
parameter $\mu^*$ may change as well. The limited evolution of
$N(E_F)/\omega_0$ suggest that $\mu^*$ should hardly change from
one type of doping to another. The average frequency $\omega_{0}$
decreases by $\sim$40 cm$^{-1}$ in the case of aluminum doping.
However, this decrease does not overcome the increase in the
parameter $\lambda$ and with $\mu^*$ in the 0.08-0.12 range, the
transition temperature T$_C$ is found to lie in between 2.8~K and
0.6~K, that is about one order of magnitude larger than the values
obtained in the case of boron doping. Even though the effect of
impurity randomness (beyond the supercell approach)
\cite{Pickett06} and possible anharmonic corrections may change
the $\lambda$ values calculated here, we expect this evolution of
the coupling constant from boron to aluminum to be robust.

We finally briefly turn to the case of hypothetical highly {\it
n}-doped phosphorus samples. In the 6.5$\%$ doping limit, the
system is degenerate, but the conduction bands and the
multi-valley structure of pristine c-Si are strongly modified,
with in particular a large splitting of the $\Gamma_{15}$ state.
We find a $\lambda$ value of 0.30, not significantly different
from the boron case and rather disappointing in view of early
arguments in favor of multi-valley semiconductors. \cite{Cohen64a}

In conclusion, we have studied the structural, electronic and
electron-phonon coupling properties in highly-doped c-Si. Our
results provide support for a standard phonon-mediated BCS-type
mechanism for the occurrence of superconductivity in B-doped
silicon. The transition temperature is expected to increase by an
order of magnitude upon aluminum doping instead of boron. It is
not clear however if the present synthesis techniques (such as the
gas immersion laser doping technique used in
Ref.~\onlinecite{Bustarret06}) may be adapted to other type of
dopants and/or to a larger doping percentage. Conflicting
phenomena, such as dopant segregation as in the case of boron
doped diamond \cite{Goss06} may significantly complicate the
search for larger T$_C$ in {\it c}-Si.

\noindent {\bf Acknowledgements:} Calculations have been performed
at the French CNRS national computing center (IDRIS, Orsay). The
authors are indebted to the French National Agency for Research
(ANR) for funding (contract ``SupraDiam"  ANR-05-BLAN-0282) and
acknowledge E. Bustarret, C. Marcenat,  and J. Boulmer for
enlightening discussions.





\begin{references}

\bibitem{Bustarret06}
E. Bustarret {\it et al.}, {\it Nature} (London) {\bf 444}, 465
(2006).

\bibitem{Cohen64a}
Marvin L. Cohen, Rev. Mod. Phys. {\bf 36}, 240 (1964);
{\it ibid}, Phys. Rev. {\bf 134}, A511 (1964).

\bibitem{Martinez85}
K.J. Chang {\it et al.}, Phys. Rev. Lett. {\bf 54}, 2375 (1985).

\bibitem{Yamanaka95}
H. Kawaji, H.-o. Horie, S. Yamanaka, and M. Ishikawa,
Phys. Rev. Lett. {\bf 74}, 1427 (1995);
K. Tanigaki {\it et al.}, {\it Nature Mater.} {\bf 2}, 653 (2003).

\bibitem{Conneta03}
D. Conn\'etable {\it et al.}, Phys. Rev. Lett. {\bf 91}, 247001 (2003).

\bibitem{Conneta01}
E. Galvani {\it et al.}, Phys. Rev. Lett. {\bf 77}, 3573 (1998);
D. Conn\'etable {\it et al.}, Phys. Rev. Lett. {\bf 87}, 206405
(2001); X. Blase, Phys. Rev. B {\bf 67}, 035211 (2003).

\bibitem{Ekimov04}
E.A.~Ekimov {\it et al.}, {\it Nature} {\bf 428}, 642 (2004);
Y.~Takano, M. Nagao, I. Sakaguchi, M. Tachiki
{\it et al.}, Appl. Phys. Lett. {\bf 85}, 2851 (2004);
E. Bustarret {\it et al.}, Phys. Rev. Lett. {\bf 93}, 237005 (2004).

\bibitem{Sacepe}
B. Sac\'ep\'e {\it et al.}, Phys. Rev. Lett. {\bf 96}, 097006
(2006).

\bibitem{Ortolani}
M.~Ortolani {\it et al.},  Phys. Rev. Lett. {\bf 97}, 097002
(2006).

\bibitem{Boeri04}
L.~Boeri, J.~Kortus, O.K.~Andersen, Phys. Rev. Lett. {\bf 93}, 237002 (2004).

\bibitem{Pickett04}
K.W.~Lee and W.E.~Pickett, Phys. Rev. Lett. {\bf 93}, 237003 (2004).

\bibitem{Ma05}
Y. Ma {\it et al.}, Phys. Rev. B 72, 014306 (2005).

\bibitem{Blase04}
X. Blase, Ch. Adessi, and D. Conn\'etable,
Phys. Rev. Lett. {\bf 93}, 237004 (2004).

\bibitem{Xiang04}
H. J. Xiang {\it et al.}, Phys. Rev. B {\bf 70}, 212504 (2004).

\bibitem{Giustino06}
F. Giustino, M.L. Cohen, S.G. Louie, Phys. Rev. Lett. (in press).


\bibitem{Baroni}
S. Baroni, A. Dal Corso, S. de Gironcoli, and P. Giannozzi,
http://www.pwscf.org.

\bibitem{PBE}
J.~P. Perdew {\it et al.} Phys. Rev. Lett. {\bf 77},  3865 (1996).


\bibitem{note1}
A Fourier-transform technique is used to extrapolate the dynamical
matrix on arbitrary ${\bf q}$-points (see
Ref.~\onlinecite{Baroni}).

\bibitem{Cardona}
For example, see: M. Chandrasekhar, H.R. Chandrasekhar, M.
Grimsditch, M. Cardona, Phys. Rev. B {\bf 22}, 4825 (1980); C.P.
Herrero and M. Stutzmann, Phys. Rev. B {\bf 38}, 12668 (1988).

\bibitem{Allen72}
Philip B. Allen, Phys. Rev. B {\bf 6}, 2577 (1972).

\bibitem{tubecase}
Such a technique has been shown to be very efficient in the case
of nanotubes. See: D. Conn{\'e}table {\it et al.}, Phys. Rev.
Lett. {\bf 94}, 015503 (2005).

\bibitem{note2}
 Without tetrahedron interpolation, a slightly different $\lambda=0.28$ value
 was found in Ref.~\onlinecite{Bustarret06} for B:Si.

\bibitem{Pickett06}
K.-W. Lee and W.E. Pickett, Phys. Rev. B {\bf 73}, 075105 (2006).


\bibitem{Goss06}
J.P. Goss and P.R. Briddon, Phys. Rev. B {\bf 73}, 085204 (2006);
E. Bourgeois {\it et al.}, Phys. Rev. B {\bf 74}, 094509 (2006).

\end{references}
\end{document}